\documentclass[cspaper]{IEEEcsmag}

\usepackage[colorlinks,urlcolor=blue,linkcolor=blue,citecolor=blue]{hyperref}

\usepackage{upmath}
\usepackage{pifont}
\usepackage{multicol}
\usepackage[square,sort,comma,numbers]{natbib}

\jvol{54}
\jnum{11}
\paper{X}
\jmonth{November}
\jname{Computer}
\pubyear{2021}

\begin{document}

\sptitle{Perspectives}
\editor{Editor: Name, xxxx@email}

\title{Toward AI Assistants That Let Designers Design}

\author{Sebastiaan De Peuter}
\affil{Department of Computer Science, Aalto University, Finland}
\author{Antti Oulasvirta}
\affil{Department of Communications and Networking, Aalto University, Finland}
\author{Samuel Kaski}
\affil{Department of Computer Science, Aalto University, Finland. Department of Computer Science, University of Manchester, UK}

\markboth{Perspectives}{Toward AI Assistants That Let Designers Design}

\begin{abstract}
AI for supporting designers needs to be rethought. It should aim to cooperate, not automate, by supporting and leveraging the creativity and problem-solving of designers. The challenge for such AI is how to infer designers' goals and then help them without being needlessly disruptive. We present AI-assisted design: a framework for creating such AI, built around generative user models which enable reasoning about designers' goals, reasoning, and capabilities.
\end{abstract}

\maketitle
\chapterinitial{Computational design} is emerging as a candidate to advance areas of design from science and engineering to the arts. Algorithmically powered tools are seeing increased adoption and are a candidate for becoming foundational to some areas of design practice, supporting thousands of professionals in their work. For example, computational methods have been developed to help structural engineers create structural elements with minimal materials usage~\cite{fairclough2019}, enabling more carbon-efficient construction. In computational drug design, there is an ongoing effort to speed up and automate the identification of promising drug molecules using computational methods~\cite{sliwoski2014}, allowing scientists to find better cures faster. And in graphic design, generative deep learning methods are helping designers create more appealing posters~\cite{guo2021vinci}.

We believe that a significant opportunity lies in AI-based computational design approaches that aim to cooperate with designers. Cooperation is better suited for the defining problem of assistance: how to infer what a designer wants without needlessly encumbering them or disrupting their creative process.

\begin{figure}
\centering
\includegraphics[width=18.5pc]{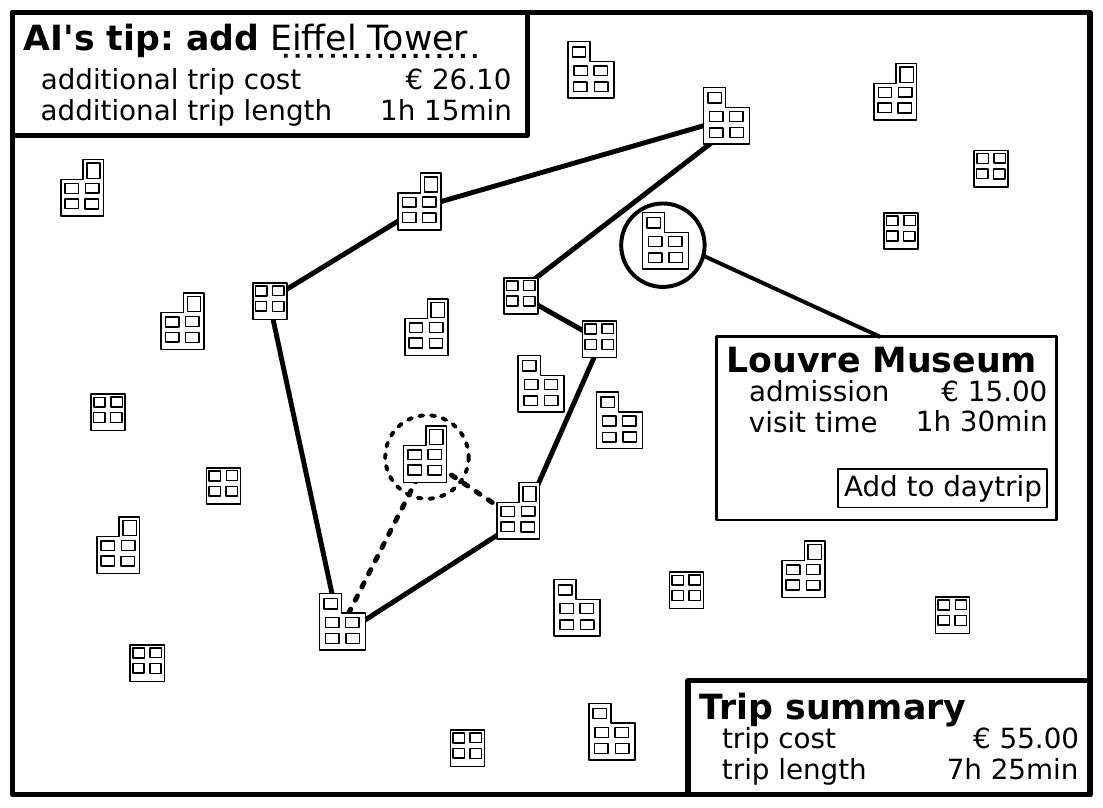}
\caption{Design problems often involve complex and tacit goals. To design a pleasant day trip, a designer must plan a round trip over a set of available points of interest (POIs). This figure shows an example user interface for this problem. A trip is constructed by selecting POIs (solid circle) and adding or removing them from the trip. We use our proposed AI-assisted design framework to implement an assistant that helps the designer by recommending to add (dashed circle) or remove a POI. Extra information related to the recommendation is shown in the top left.}
\label{fig:tripplanning}
\end{figure}

Consider the apparently simple problem of designing a pleasant day trip when visiting a city abroad. We will use this as a running example throughout this paper. The goal is to select a set of points of interest (POIs) which form a maximally enjoyable trip (\textbf{Figure~\ref{fig:tripplanning}}). To help us with this task we may choose to use an optimization algorithm. However, this would require that we specify our idea of an ideal day trip as an objective function to a computer, which we would quickly find to be challenging. Knowing what we want is difficult before we explore what is available. How many museums are we willing to visit in a day? Would more museums be okay if we planned some shopping in between? Or should we break up the museum visits with some sightseeing and do the shopping at the end of the day?

This day trip planning problem captures a defining characteristic of problem-solving in design: it may not be clear from the outset what the options or the objectives are, rather these are learned by exploration and trial and error. The problem is under-specified. Designers across fields engage in iteratively constructing solution candidates and develop their design goals during this process. Design goals that were initially tacit, evolve and concretize as designers work~\cite{dorst2001}. Any method that aims to support designers should appreciate this.

Currently most design tools separate the issue of communicating goals from the issue of solving the design problem. First determining the goal and then solving for it allows one to create solvers that work independently from the designer. This follows a general preference for developing systems that work autonomously, even if in practice they do not~\cite{dafoe2021}. This approach therefore relies crucially on having good goal communication; bad communication will lead to the solver solving for the wrong goal. For a designer it is difficult to get the goal right because it usually needs to be communicated separately from working with the design. This approach simply does not adequately appreciate the explorative process by which designers refine their goals.

Unfortunately, few adequate AI methods exist that can support designers in communicating their goals. The most well-studied approaches are inverse reinforcement learning~(IRL) and elicitation methods (see box). IRL learns goals from demonstrations but is prone to error if the cognitive bounds that determine what a human can demonstrate are not modeled correctly. Elicitation methods elicit a goal by asking questions about it. However, it can take hundreds of questions to elicit a non-trivial goal.

\begin{figure*}
  \centerline{\includegraphics[width=26pc]{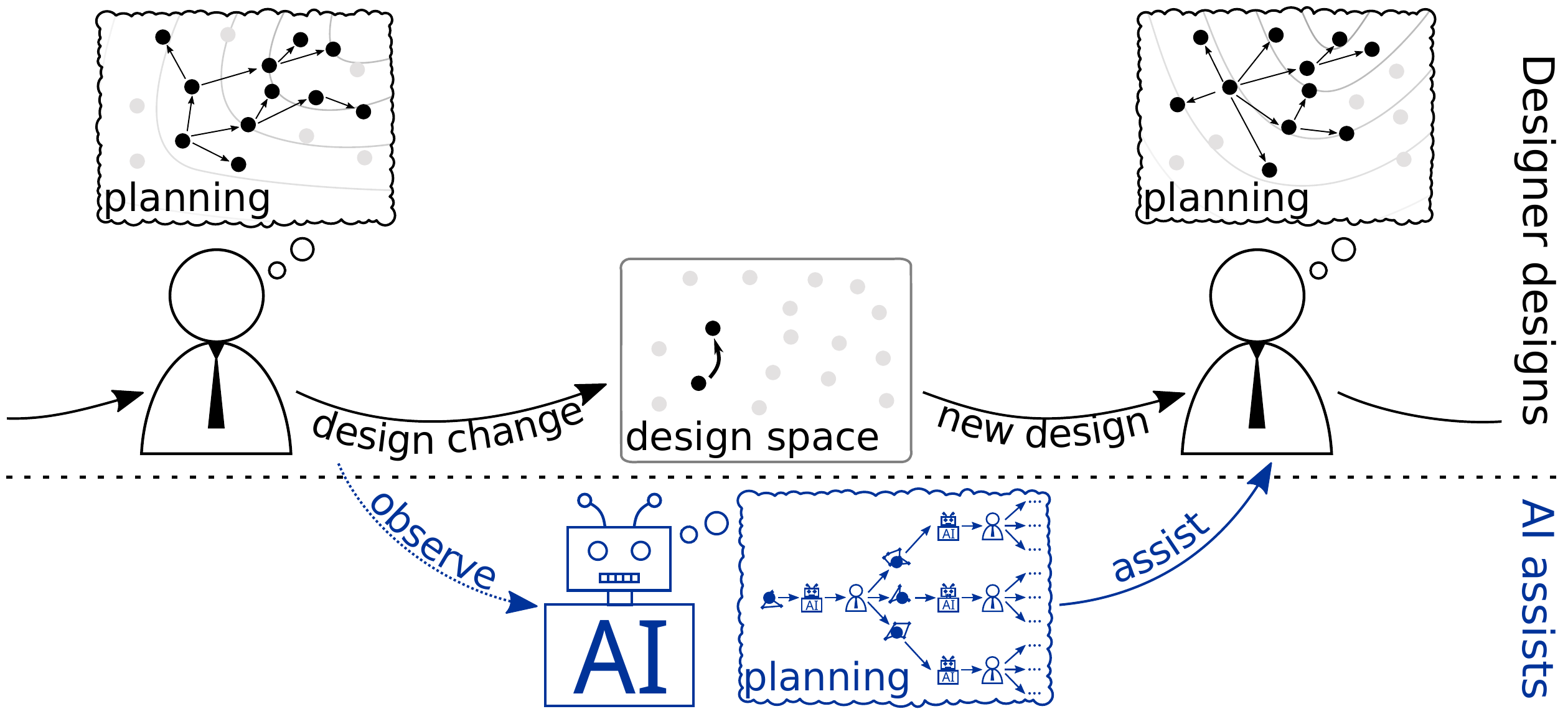}}
  \caption{To help designers, AI should appreciate the explorative and evolving character of their thinking. Designers generate solutions not only to solve a problem but also to learn about it, including its objectives and constraints. Within a design process a designer can be seen as mentally planning over a design space based on a utility function (shown as contours). The utility function evolves as design progresses. The AI should cooperate in this creative process, for example by proposing high-quality solutions, and complement the designer’s problem-solving. To do so it needs to know the designer’s utility function. We propose to create AI assistants (shown in blue) that can infer this utility from observations, and then use it to assist a designer.}
  \label{fig:AIAD}
\end{figure*}

\subsection{Cooperative design assistance}
There is an urgent need for developing AI methods designed to cooperate with designers. Such methods can be thought of as assistants that communicate with a designer about the design goal while at the same time supporting them in working toward that goal. Intuitively, these assistants plan changes to the design, just like designers do but faster, based on what they think the designer wants and convey the results back to the designer.

The designer remains the primary actor in the design process but is empowered in their decision making by the assistant. This allows the designer to work more effectively, making their behavior more indicative of the goal they are working toward. Keeping the designer in the loop also reduces our reliance on goal communication, as any error in the communicated goal can be corrected immediately. As an active participant the designer is free to explore and try things out in order to refine their goals. Further, cooperation can use the designer's creative abilities and expertise to the fullest extent, and serves to reduce frustration due to a lack of control.

In this paper we propose \emph{AI-assisted design}, a general-purpose framework for cooperative assistants in design problems. Cooperative design assistance has been proposed for specific design problems~\cite{guo2021vinci}, but no general framework currently exists. The framework is designed to support a wide range of interactions between assistant and designer, though our focus here will be on design change recommendations. Cooperation requires socially intelligent assistants that understand and can communicate with designers~\cite{dafoe2021}. AI-assisted design incorporates the most important aspects of cooperation: understanding and accounting for designers' goals, reasoning, and capabilities. It uses a \emph{generative user model}, a model of how goals result in the behavior we observe from designers, to infer a designer's goal from their behavior and to plan how to best assist the designer.

\section{AI THAT PLANS LIKE DESIGNERS DO}
Rethinking cooperative AI for design needs to start from the way design is conceptualized. Design is usually thought of as an optimization problem, where an ideal point must be identified within a parameter space~\cite{rao2019}. In our view it is better to formalize it as a decision process, where designs are states and \emph{design changes} are actions. The goal is encoded as a \emph{utility function} which the decision process seeks to maximize. In the day trip example, every visit added to the trip is a decision that is part of the decision process that leads to a final trip.

This approach is well-justified. Designers are decision makers: they reason about the changes they can make to a design and choose one (\textbf{Figure~\ref{fig:AIAD}}). Say we were planning a trip and needed to choose between visiting the Louvre Museum or the Eiffel Tower. We would consider both options and any potential future changes -- if we add the Eiffel Tower it will be easy to visit the Quai Branly Museum next as it is close by -- and choose what we like most. Integrating a human decision maker into an optimization process is difficult because humans do not think or work like optimizers do. However, if design is formalized as a decision process, there is a clear opportunity to create cooperation around the design decisions that have to be made.

Within this paper we will assume that a general parameterized design space is given, and will focus solely on the decision making aspect of design. In established design domains, standardized parameterizations of the design space often exist. In the trip planning example a design can simply be represented as a set of POIs. We conceptualize the result of putting the design into practice as \emph{outcomes}. In the day-trip example, the time spent walking is an outcome.

\begin{figure*}
\fbox{\begin{minipage}{\textwidth}
\section{BOX: PROGRESS ON COMPUTATIONAL METHODS FOR DESIGN ASSISTANCE}
Several intelligent interaction techniques have been proposed for computational design tools, from interactive example galleries~\cite{lee2010} to smart sliders~\cite{igarashi2007}. However, at the end of the day, they are all based on some computational method of (1) inferring what the user wants (goal communication) and (2) choosing what to present and how. Presently available methods in AI are unable to make useful suggestions without assuming one of the following: (1) very large training sets (for supervised learning), (2) lots of training feedback from the designer, (3) stationary preferences, or (4) a designer who can demonstrate exactly what they want.
\begin{itemize}
\item In \emph{trial and error optimization}~\cite{meignan2015} a designer uses a solver directly without support for communicating their goal. This approach relies on giving the designer multiple attempts to specify their goals to get the description right. This seldom results in a good description, causing the solver to solve for a goal that differs from what the designer intended.
\item In \emph{interactive optimization}~\cite{meignan2015}, a designer steers the optimization process by selecting samples that are promising for further computation ("I like this design, give me more like that"), or by providing reference points to objectives ("I want this objective to reach at least 3.0 and this one 4.5 points"). This type of steering is an easier form of goal communication but is also less informative.
\item Elicitation methods, which form a subclass of interactive optimization, try to elicit a goal by asking questions about it. \emph{Preference learning}, for example, elicits a goal by asking a designer about their preferences, usually by asking them to indicate their preferences over pairs of candidate solutions. A partial~\cite{deb2014} or complete~\cite{gonzalez2017} utility function which encodes the goal is inferred from the answers, and a solver is used to produce a final design. Because this type of communication is less informative it can take a prohibitively large number of comparisons to learn a complex utility function. Moreover, these methods generally assume a stationary utility function, which is problematic if designers may develop their goals over time.
\end{itemize}

Other AI methods have been proposed to tackle goal communication and interactive decision making in general decision problems. In principle, our definition of design as a decision process allows us to apply these to design problems.
\begin{itemize}
\item \emph{Inverse reinforcement learning}~(IRL)~\cite{abbeel2004} learns a goal based on one or more demonstrations. A problem in IRL is that as bounded agents, humans are limited in what they can demonstrate. Knowing these bounds is essential for correctly interpreting behavior, but inferring them together with goals from demonstrations is often impossible~\cite{mindermann2018}. Therefore, the problematic assumption is often made that the demonstrator has no bounds and gives demonstrations of exactly what they want.
\item \emph{Cooperative inverse reinforcement learning}~(CIRL)~\cite{hadfield2016} is a cooperative assistance method that introduces an assistant that works side-by-side with a human. CIRL is similar to the framework we propose but gives the assistant complete autonomy. This means that the human has to keep careful watch of the assistant to correct any mistakes it makes.
\item In \emph{shared autonomy}~\cite{javdani2015}, another cooperative assistance method, the human directs an assistant as it solves a decision problem. However, the assistant is given significant freedom in how it acts based on these directions, making the human's control over the assistant limited.
\end{itemize}
\end{minipage}}
\end{figure*}

\section{AI-ASSISTED DESIGN}
The ability to support designers cooperatively would be a game-changer. However, this requires knowing the designer’s goal. In what follows we propose an AI assistant that helps a designer based on an inferred utility; the utility is inferred \emph{while giving assistance} (blue part of Figure~\ref{fig:AIAD}). We focus here on assistance in the form of design change recommendations, and generalize to other forms of interaction later.

The assistant recommends design changes. This filters the large set of available design changes to just those that the assistant thinks are interesting and worth considering for the designer. For example, in trip planning the assistant may determine that we are most interested in visiting fine arts museums, and thus will recommend fine arts museums that can easily be incorporated into our trip. For the designer this reduces the complexity of designing to choosing the best option out of a limited set of recommendations.

Unlike regular recommendation engines that pick interesting objects, recommendations here are \emph{what-if} scenarios. For every recommended design change the resulting design is shown together with additional relevant information, in particular the outcomes. This supports the counterfactual thinking (i.e. hypothesizing about the effects of design changes) that designers do while designing. In trip planning a recommendation to visit the Eiffel Tower can immediately show what the resulting trip looks like on a map and what the additional trip length and cost would be (Figure~\ref{fig:tripplanning}). As outcomes can be too complex to be accurately predicted by a designer, showing them immediately takes a lot of the guesswork out of choosing design changes. The designer can either accept one of the recommendations or choose a design change themselves. This is observed by the assistant and used to adapt the inferred utility function. The recommendations that the assistant makes change at every iteration, in response to changes in the design and the assistant's evolving inferences about the designer's utility.

The assistant uses its interaction with the designer to infer the utility. Recommendations are a good baseline choice for this interaction but this framework can accommodate other interactions as well. Richer forms of interaction will be more informative for the assistant and can further empower the designer. For instance, simple questions such as "How much are you willing to spend on this day trip?" can be extremely informative yet easy to answer.

\section{CREATING A COOPERATIVE ASSISTANT}
Successful cooperation requires two capabilities: that the assistant (i) can estimate the designer's utility and -- given that utility -- (ii) can predict how the designer will behave in future situations~\cite{dafoe2021}. The first ensures that the assistant can act in the interest of the designer, for example by making recommendations that improve the design. The second allows the assistant to plan its assistance based on how the designer will react to it. For instance, as humans are bounded agents with incomplete knowledge, a designer may not see the value of a recommendation, even if it is optimal~\cite{elmalech2015}. The assistant can mitigate this by tailoring its recommendations to be understandable to the designer.

\subsection{Modeling designers}
To allow the assistant to act collaboratively we propose to equip it with a \emph{generative user model}, that is, a generative model of how human reasoning translates an internal utility function into behavior. This allows the assistant to evaluate possible actions by forward-simulating their effect on the user. We emphasize the contrast here to earlier work on user modeling which focused on modeling knowledge and preferences for personalization~\cite{kobsa2001}, as opposed to a designer’s thinking.

The user model should capture the most prominent factors (e.g. cognitive bounds) that influence human behavior. By parameterizing these factors they can be inferred from observations, together with the utility function, up to limitations on identifiability~\cite{mindermann2018}. If the user model correctly captures these factors, their effect on the designer's behavior can be separated from that of the utility. This can allow an assistant to infer the utility even from highly biased behavior.

This generative user model allows the assistant to perform the two types of reasoning necessary for cooperation. (i) Inference of the user model parameters and the utility function allows the assistant to determine and act in the designer's interest. (ii) Once the utility has been inferred, the assistant can simulate the user model to predict its effect on the designer's behavior, and plan its assistance accordingly.

So far, we have not made assumptions about the internals of the user model. The only requirements are that the model is generative and models the behavior for all interactions supported by the assistant. Of course, creating highly accurate comprehensive models of a designer's thinking is presently out of reach. But this need not be the goal; models do not have to be perfect to be useful. Other interactive decision making approaches have shown success using relatively simple models of human behavior~\cite{reddy2018}.

Under these requirements, computational rationality~\cite{lewis2014} is emerging as one promising approach for building these user models. Computational rationality is a cognitive theory that postulates that outwardly irrational human behavior corresponds to rational utility-maximizing behavior under a set of bounds and a subjective model of the problem. The utility being maximized here is the internal utility of the designer. The subjective model of the task corresponds to the designer's own understanding of the design space. Computational rationality is a well-founded theory of human behavior and is built on a decision-theoretic view of human reasoning, making it well-suited for how our framework formalizes design. Therefore, these cognitive bounds are an ideal way to introduce prior knowledge into the user model, reducing the need for large amounts of training data, yet still producing flexible models.

\subsection{Planning assistive actions}
The assistant \emph{plans}, sequentially, its interactions with the designer, taking into account how useful they will be to the designer and how much the resulting observations will improve future assistance. The interactions with the designer form a decision process, different from the designer's decision process but with the same goal: to maximize the utility of the design produced.

The assistant plans over this decision process to find a policy for interacting with the designer. The user model supports this planning by allowing the assistant to evaluate how the designer will change the design in response to interactions. Interactions have a short and long term effect on the design process. They may lead to an immediate increase in utility, for example by suggesting good design changes. In the long term they will improve assistance because the resulting observations can be used to better infer the utility. Planning must consider both of these effects.

\section{AI-ASSISTED DAY TRIP PLANNING}
We demonstrate our framework on the trip planning example from Figure~\ref{fig:tripplanning}. A designer must plan a maximally enjoyable day trip lasting no more than 12 hours in a city consisting of 100 points of interest (POIs). Choosing the order in which the chosen POIs are visited is a traveling salesperson problem (TSP) and is done automatically. At every iteration of the design process the assistant makes a single recommendation for adding or removing a POI. The utility is a weighted combination of two scores, one scoring the monetary cost of the day trip and another measuring enjoyment based on the time spent at POIs and the time spent walking.

Note that while this specific design problem could in principle be solved using a (sequential) recommendation system, this is not the case in general. As the user base of travellers is large, it would be possible to collect a data base needed for training a recommendation system. However, most other design tasks have significantly smaller user bases. For instance, there are not enough pedestrian suspension bridges being engineered to train a recommendation system for that application.

The implementation works as follows. At every iteration the assistant makes a single recommendation for a change to the trip. For example in Figure~\ref{fig:tripplanning} it recommends to add the Eiffel Tower. A (simulated) designer then makes a change to the trip, taking this recommendation into account. In Figure~\ref{fig:tripplanning} they chose to add the Louvre museum. This change is applied and a new trip is automatically produced, with the POIs ordered to minimize the travel distance. The assistant observes the designer's choice and uses it to update its posterior over user models and utilities. This posterior is represented using a weighted particle set. In the next iteration the assistant will then plan over this particle set to determine what design change it should recommend next.

The user model we have constructed for this demonstration is based on cognitive theories of noisy decision making. In our model the designer chooses a design change in three phases: first the designer chooses a design change from all available changes, next they consider the assistant's recommendation and choose whether to switch to it. Last, if the chosen design change is worse than doing nothing, the designer chooses to do nothing. These three choices are made in a Boltzmann rational way (i.e. the log-probability of a choice is proportional to its utility). The utility of the available design changes is calculated using limited horizon planning. This is motivated by the idea that people think a few steps into the future when weighing their options.

\begin{figure}
\centering
\includegraphics[width=18.5pc]{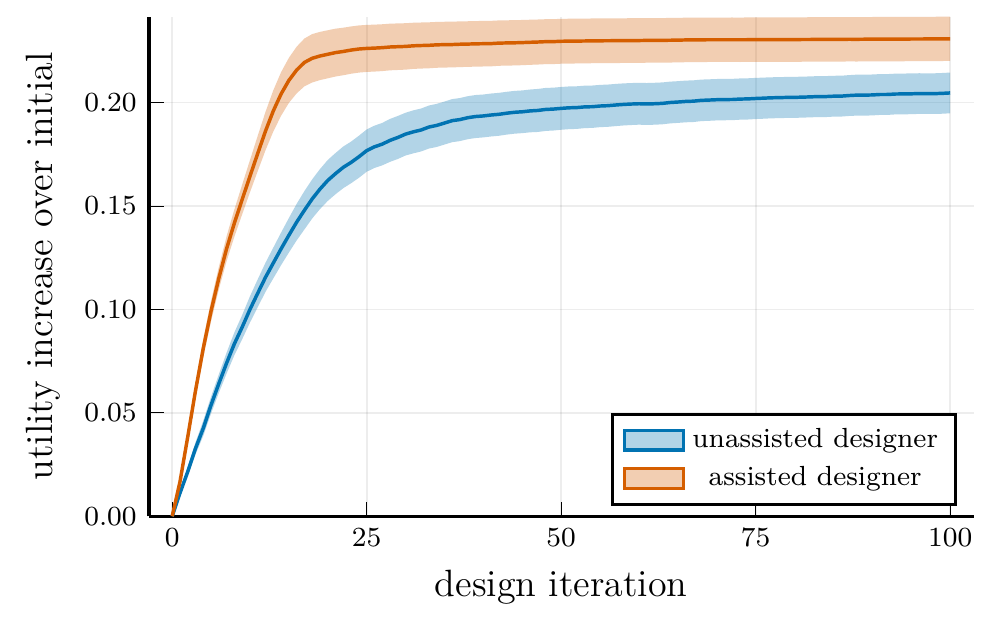}
\caption{We compare how quickly simulated designers improve the design of a day trip on average. Unassisted designers worked independently while assisted designers received recommendations from an AI assistant implemented using our framework. We see that assisted designers improved their day trips significantly faster and generated significantly better designs. The shading shows twice the standard error, calculated over multiple randomly sampled designers and utilities.}
\label{fig:AIAD_performance}
\end{figure}

Reasoning about the effect of design changes involves reasoning about the resulting optimal order for the POIs. Here our user model is inspired by prior research on how humans solve TSP problems, which suggests they work iteratively, using visual heuristics like the maximum angle insertion criterion~\cite{macgregor2000}. When a POI is added to a trip, the user model places it wherever it creates the largest angle with existing POIs. When removing a POI from a trip the order remains unchanged.

\textbf{Figure~\ref{fig:AIAD_performance}} shows the improvement in trip utility over an initial empty trip across subsequent design iterations for assisted and unassisted simulated designers. We see that within the 100 design iterations considered, the assistant helps the designer achieve a significantly better design. More importantly, we see that assisted designers improve their designs significantly faster than designers without assistance. Every designer was a randomly sampled user model from a hypothesis set known to the assistant. This allows us access to the internal utility function of these designers so that we can evaluate the true utility of the trips produced. We ran 200 experiments with randomly sampled user models of both the assisted and unassisted designer. The POIs were randomly generated.

\section{A CALL TO ACTION}
Cooperative assistants that facilitate goal communication while at the same time helping a designer could revolutionize computational design. Current two-step approaches, which first infer the goal and then solve for it, are fundamentally limited by the quality of goal communication. Cooperative approaches can keep the designer in the loop as an active participant and resolve miscommunications quickly. A designer who is supported by an AI assistant is also able to better show what they want, leading to better goal communication overall. Making the designer an active participant is better suited to the trial and error process designers use to develop their conception of the design goal, and can make better use of their expertise and creativity. Cooperative assistants could therefore make for much better design tools than are available currently, increasing productivity in design problems that are central to much of today's research and development.

In this paper we have introduced AI-assisted design, a framework for developing cooperative assistants for design problems. We now highlight four open challenges for cooperation in design:\\
\textbf{The assistant should support a multitude of interactions.} We have introduced recommendations here as a general type of interaction but other more specialized forms may be more suitable for individual design problems. To offer the best assistance it can, the assistant should have multiple types of interaction at its disposal. Interactions should be developed to be maximally useful to designers and to be informative to the assistant.\\
\textbf{The assistant must plan its interactions effectively.} With a large set of interactions at its disposal the assistant must choose how to interact. A trade-off needs to be made between interactions that are most useful now and interactions that will improve the assistant's understanding of the designer and thus improve assistance in the future.\\
\textbf{Cooperation with the designer requires a sufficiently accurate user model.} To cooperate, the assistant needs a good model of the designer's behavior. AI-assisted design does not commit to a single modeling paradigm, but we have argued that computational rationality is an especially promising option for introducing scientifically grounded prior knowledge.\\
\textbf{New machine learning methods are needed to support planning and inference with the user model.} Inferring the utility and the parameters of the user model needs to be done in a data-efficient way to minimize unnecessary interactions. This will require a combination of effective interaction planning and novel machine learning methods. The multi-agent nature of cooperation requires complex nested reasoning within the user model, making forward simulation computationally expensive. Creating assistants that can swiftly provide helpful assistance therefore requires new machine learning methods and approximations to ensure that interaction planning can be done in real time.

\section{ACKNOWLEDGMENT}
We would like to thank Julien Gori, Andrew Howes, Jussi Jokinen, and Pierre-Alexandre Murena for their valuable feedback and suggestions. This work was supported by the Academy of Finland (flagship programme: Finnish Center for Artificial Intelligence, FCAI; grants 328400, 319264, 292334). We acknowledge the computational resources provided by the Aalto Science-IT project.

\bibliographystyle{IEEEtranN}
\bibliography{references}

\begin{IEEEbiography}{Sebastiaan De Peuter}{\,}is a doctoral candidate at Aalto University, Finland. Contact him at sebastiaan.depeuter@aalto.fi.
\end{IEEEbiography}

\begin{IEEEbiography}{Antti Oulasvirta}{\,}is a computational cognitive scientist and associate professor at Aalto University, Finland. He leads the interactive AI research programme at the Finnish Center for AI.  Contact him at antti.oulasvirta@aalto.fi.
\end{IEEEbiography}

\begin{IEEEbiography}{Samuel Kaski}{\,}is a professor of Computer Science at Aalto University and professor of AI at the University of Manchester. He leads the Finnish Center for Artificial Intelligence FCAI and ELLIS Unit Helsinki. Contact him at samuel.kaski@aalto.fi.
\end{IEEEbiography}

\end{document}